# Phylogenetic analyses of the severe acute respiratory syndrome coronavirus 2 reflected the several routes of introduction to Taiwan, the United States, and Japan

Tomoko Matsuda[1], Hikoyu Suzuki[1], Norichika Ogata[1,2,*]
[1]Nihon BioData Corporation, Kawasaki, Japan,
[2]Medicale Meccanica Inc., Kawasaki, Japan
*norichik@nbiodata.com

*Abstract*— **Worldwide Severe acute respiratory syndrome coronavirus 2 (SARS-CoV-2) infection is disrupting in the economy and anxiety of people. The public anxiety has increased the psychological burden on government and healthcare professionals, resulting in a government worker suicide in Japan. The terrified people are asking the government for border measures. However, are border measures possible for this virus? By analysing 48 almost complete virus genome sequences, we found out that the viruses that invaded Taiwan, the United States, and Japan were introduced independently. We identified thirteen parsimony-informative sites and three groups (CTC, TCC, and TCT). Viruses found outside China did not form a monophyletic clade, opposite to previous study. These results suggest the difficulty of implementing effective border measures against this virus.**

## I. Introduction

The outbreak of a new coronavirus (SARS-CoV-2) [1] has had a significant impact on neighbouring countries with transportation to and from China due to the spread of SARS-CoV-2. It is reasonable to think that restricting human movement is effective in managing viruses, as human movement accelerates the spread of the virus. A total ban on traffic could have helped control virus transmission [2]. Movement of people across the sea by plane or ship is considered necessary in the movement of viruses [3]. Several governments announced strengthened border measures to prevent viral border invasion [4, 5]. The various restrictions associated with the pandemic hurt not only the economy [6], but also the mental health [7, 8]. The fear of the virus has resulted in discrimination and excessive demands on the government. A 37-year-old government worker who worked on antivirus committed suicide in Japan [7]. With the availability of viral genome sequences, research is being conducted into many different aspects [9-13]. An analysis of previously collected viral genomes indicated that viruses outside China formed a monophyletic clade, suggesting the effectiveness of border measures [14]. Recently, more complete viral genome sequences have been published, providing more resources to discuss virus transmission [15]. Although several phylogenetic trees of the virus have been published, many analyzes including fragment sequences did not provide accurate phylogenetic information. Therefore, we scrutinized the available viral sequences and collected only full-length viral genomic sequences for phylogenetic analysis.

## II. Materials and Methods

<u>*Data Mining*</u>

The genome sequences of SARS-CoV-2 isolated from human and “BatCoV RaTG13” (Accession No. GWHABKP00000000), which is most closely related to SATS-CoV-2 [9], were obtained from 2019 Novel Coronavirus Resource [15] (Accessed 22 Feb, 2020). We choose 51 longer sequences. The shortest one is 26,973 bps and the longest one is 29,903 bps. One of the longest sequences, “Wuhan-Hu-1” (Accession No. MN908947), was used as the reference genome. We also obtained the genome sequence of “Bat SARS-like coronavirus” (Accession No. MG772933) from GenBank as the outgroup according to the previous study [9]. In total, we used 49 SARS-CoV-2 sequences (Table 1) and 2 bat coronavirus sequences (Table 2)

<u>*Isolation of open reading frames*</u>

The open reading frames (ORF) information of the reference genome were obtained from GenBank. We used the translated ORF amino-acid sequences as the queries, and searched other SARS-CoV-2 genomes for ORF by using FATE [16] with TBLASTN engine and other default parameters. The ORF sequences were cut out from the genome sequences based on the results of FATE by using BEDTools [17].

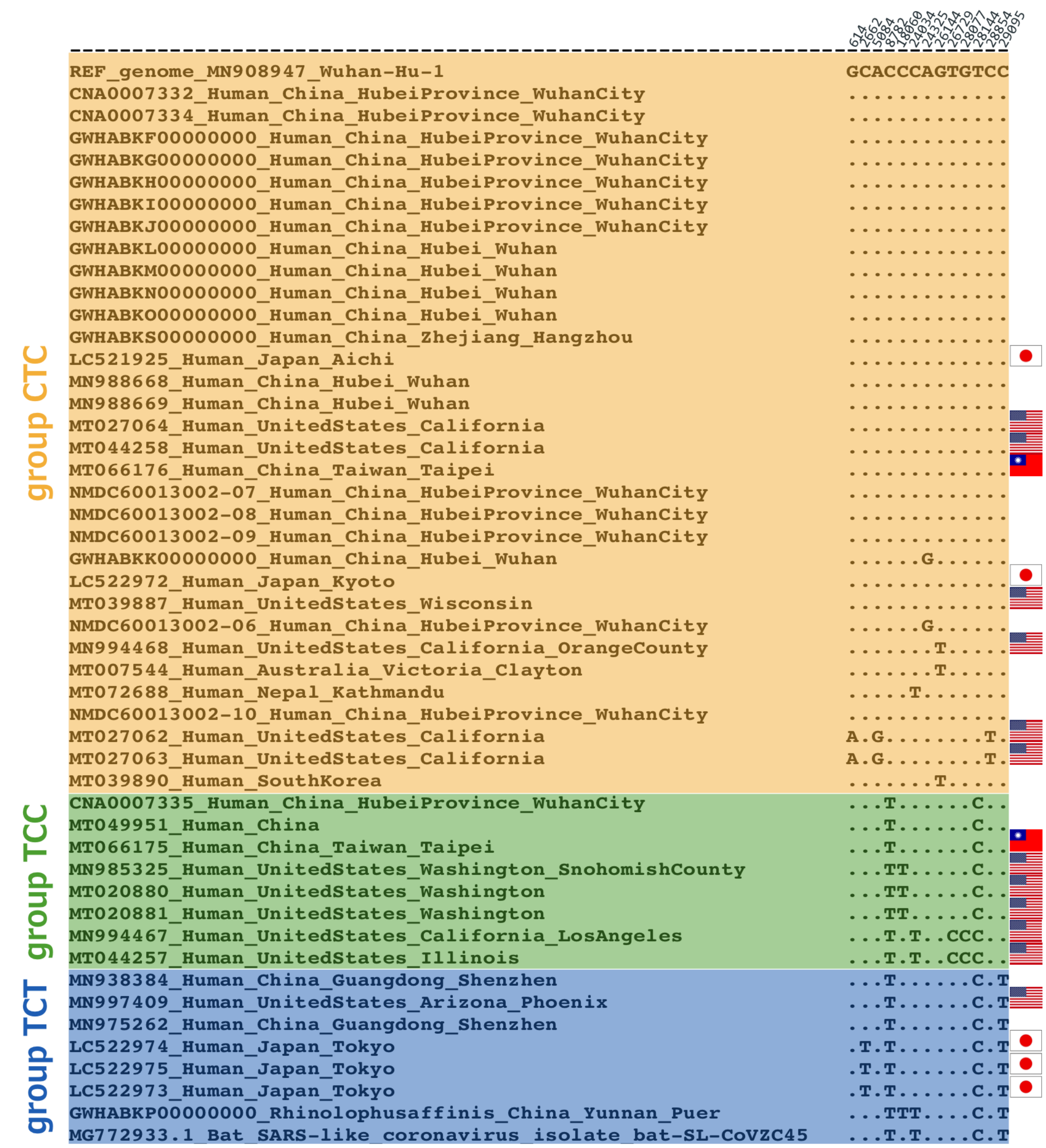

**Figure 1. Parsimony-informative sites found from SARS-Cov-2 ORF sequences**

There were 13 parsimony-informative sites in 29,145 bps alignment of sequences from 47 samples isolated from human. Dots indicate the same as the reference base on the top of the alignment. A, C, G or T indicate the difference in the reference base. Blue, green, and yellow highlight indicate group TCT, TCC, and CTC respectively. Group TCT was defined by T8782, C28144 and T29095. Group TCC was defined by T8782, C28144 and C29095. Group CTC was defined by C8782, T28144 and C29095. All sites including gaps were removed from the analysis. Bases of the same sites on the 2 coronavirus genomes isolated from bat were shown in bottom and 2nd line from the bottom of the alignment, respectively.

*Codon-based multiple sequence alignment and extraction of variable/parsimony-informative sites*

Isolated ORF sequences were aligned for each gene based on codons by using DIALIGN-TX [18]. The largest gene “orf1ab” was divided into upstream and downstream regions and aligned respectively because this gene has the ribosomal slippage site. We removed three samples, “Human-UnitedStates-Illinois-Chicago” (Accession No. MN988713), 'Human-Finland-Lapland' (Accession No. MT020781), and “Human-China-HubeiProvince-WuhanCity” (Accession No. NMDC60013002-05) from the analysis because their sequences include mixed bases. Then, we extracted variable/parsimony-informative sites based on the multiple sequence alignment by using in-house Perl script.

*Phylogenetic analysis*

Maximum-likelihood phylogenetic analyses based on the multiple sequence alignment of ORF were performed under the seven conditions below, (1) DNA-based no separate, (2) separating 1st/2nd and 3rd codon position, (3) separating each codon position, (4) separating each gene, (5) separating each gene and also 1st/2nd and 3rd codon position, (6) separating each gene and each codon position, and (7) protein-based separating each gene. For each condition, the best evolutionary models were estimated by using ModelTest-NG [19] with default parameters, followed by estimating the phylogenies of SATS-CoV-2 by using RAxML-NG [20] with '--all' option and the best Bayesian information criterion (BIC)-supported evolutionary models. Under any conditions, the outgroup was set to 'Bat SARS-like coronavirus' and the number of bootstraps replicates was set to 1000. Obtained phylogenies were colored by using MEGA X [21]. All analyses data and in-house scripts are available [22].

## III. Results and Discussion

60 variable sites were found from 29,145 bps alignment of SARS-CoV-2 ORF sequences from 47 samples isolated from human (Figure S1). There were 13 parsimony-informative sites (614, 2662, 5084, 8782, 18060, 24034, 24325, 26144, 26729, 28077, 28144, 28854 and 29095-th nucleotide of referential SARS-CoV-2 genome (MN908947)), of which 7 were non-synonymous (Table 3). There were 3 important parsimony-informative sites. The first important parsimony-informative site was 8782-nd nucleotide of referential SARS-CoV-2 genome (MN908947). Referential nucleotide is C and alternative nucleotide is T. The second important parsimony-informative site was 28144-th nucleotide of referential SARS-CoV-2 genome (MN908947). Referential nucleotide is T and alternative nucleotide is C. This substitution results in changes in amino acids 84 of ORF8, a previously indicated polypeptide involved in promoting virus transition from bat to human [24, 25]. The third important parsimony-informative site was 29095-th nucleotide of referential SARS-CoV-2 genome (MN908947). Referential nucleotide is C and alternative nucleotide is T. Therefore, we named 3 groups, TCC, TCT and others (CTC) (Figure 1). Viruses found outside China did not form a single group, opposite to previous study [14]. This result suggests that the viral were introduced to each country several times.

Phylogenies of SARS-CoV-2 were constructed based on the ORF sequences (Figure 2 and Figure S2). DNA-based phylogeny under the separating each codon position condition (Figure 2) showed the highest final log-likelihood and 2nd-lowest AIC score among DNA-based phylogenetic trees. On the other hand, protein-based phylogenetic tree (Figure S2F) might not be reliable because the tree was constructed based on less informative sites except for the synonymous substitution sites. According to our phylogeny, the group CTC formed a monophyletic clade, including all samples isolated from Chinese in Wuhan except for "CNA0007335-Human-China-HubeiProvince-WuhanCity".
Samples of the group TCC and TCT were paraphyletic for group CTC and located near the root of the tree. Our phylogenies suggest that the "Bat-SARS-like coronavirus" (MG772933) and "BatCoV RaTG13" are more closely related to samples included in the group TCC and TCT than samples included in the group CTC. Some of the phylogenies constructed under other conditions suggest the different phylogenetic position of group TCC and TCT, however, support the monophyly and the same phylogenetic position of group CTC (Figure S2). These results indicate that SARS-CoV-2 widely spread in Wuhan is different from that of other areas. In areas other than Wuhan, where virus invasion has been observed, the damage has not seemed to be as severe as Wuhan, however the spread of group CTC might increase the damage in the near future. From the perspective of virus control, sequencing should be performed since viruses have multiple properties that is hard to be distinguished by qPCR.

The viral genome sequences "MT039890-Human-Sourth-Korea" and "GWHABKG0000000-Human-China-Hubei-Province-Wuhan-City" have many sample-specific substitutions in their sequence. The viral genome sequence "MN988713-Human-USA-Illinois-Chicago" has several ambiguous bases in parsimony-informative sites, discovered in this study. Interestingly, the ambiguous bases can be explained by the mixture of group TCC and CTC (Table 4). This fact suggests that "MN988713-Human-USA-Illinois-Chicago" contains at least two different viral genomes. We could corroborate this hypothesis after next generation sequence data release.

To determine whether viral sequences have novel mutations acquired during spreading the infection or just reflect the diversity of the place of origin, various SARS-CoV-2 genome sequences should be collected in the place of origin.

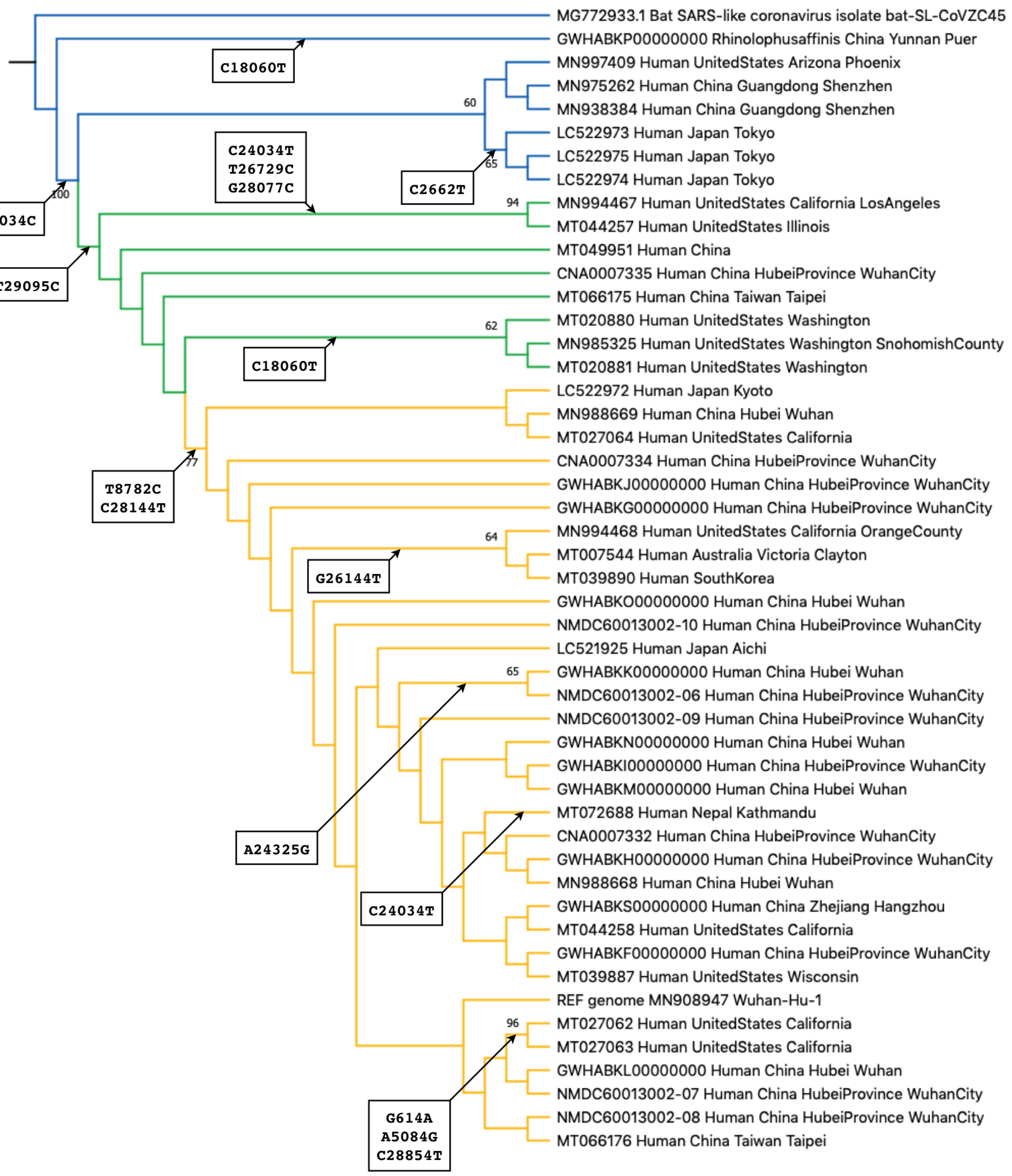

**Figure 2. Topological phylogeny of SARS-CoV-2 estimated with maximum-likelihood**

A maximum-likelihood phylogenetic analysis based on the multiple sequence alignment of ORF were performed under the condition of separating each codon position. Blue, green, and yellow branches indicate group TCT, TCC, and CTC respectively. Substitutions on the parsimony-informative sites were mapped to the corresponding branches. SARS-CoV-2 isolated from human formed a monophyletic clade with 100% bootstrap value. Group CTC formed a monophyletic clade with 74% bootstrap value. Group TCC was paraphyletic respect to the group CTC. Group TCT was paraphyletic respect to the clade consists of group CTC and TCT. Final log-likelihood was -55794.362675. AIC score was 112649.332493. Branch length was meaningless.

**Table 1**. Names and Length of SRAS-CoV-2 genome sequences used in this study.

| # | Genome Sequences | Length (nt) | Database | Acc. Number |
|---|---|---|---|---|
| 1 | CNA0007332_Human_China_Hubei_Province_Wuhan_City | 29866 | CNGBdb | CNA0007332 |
| 2 | CNA0007334_Human_China_Hubei_Province_Wuhan_City | 29868 | CNGBdb | CNA0007334 |
| 3 | CNA0007335_Human_China_Hubei_Province_Wuhan_City | 29872 | CNGBdb | CNA0007335 |
| 4 | GWHABKF00000000_Human_China_Hubei_Province_Wuhan_City | 29899 | Genome Warehouse | GWHABKF00000000 |
| 5 | GWHABKG00000000_Human_China_Hubei_Province_Wuhan_City | 29889 | Genome Warehouse | GWHABKG00000000 |
| 6 | GWHABKH00000000_Human_China_Hubei_Province_Wuhan_City | 29899 | Genome Warehouse | GWHABKH00000000 |
| 7 | GWHABKI00000000_Human_China_Hubei_Province_Wuhan_City | 29890 | Genome Warehouse | GWHABKI00000000 |
| 8 | GWHABKJ00000000_Human_China_Hubei_Province_Wuhan_City | 29883 | Genome Warehouse | GWHABKJ00000000 |
| 9 | GWHABKK00000000_Human_China__Hubei_Wuhan | 29825 | Genome Warehouse | GWHABKK00000000 |
| 10 | GWHABKL00000000_Human_China_Hubei_Wuhan | 29891 | Genome Warehouse | GWHABKL00000000 |
| 11 | GWHABKM00000000_Human_China_Hubei_Wuhan | 29852 | Genome Warehouse | GWHABKM00000000 |
| 12 | GWHABKN00000000_Human_China_Hubei_Wuhan | 29854 | Genome Warehouse | GWHABKN00000000 |
| 13 | GWHABKO00000000_Human_China_Hubei_Wuhan | 29857 | Genome Warehouse | GWHABKO00000000 |
| 14 | GWHABKS00000000_Human_China_Zhejiang_Hangzhou | 29833 | Genome Warehouse | GWHABKS00000000 |
| 15 | LC521925_Human_Japan_Aichi | 29848 | GenBank | LC521925 |
| 16 | LC522972_Human_Japan_Kyoto | 29878 | GenBank | LC522972 |
| 17 | LC522973_Human_Japan_Tokyo | 29878 | GenBank | LC522973 |
| 18 | LC522974_Human_Japan_Tokyo | 29878 | GenBank | LC522974 |
| 19 | LC522975_Human_Japan_Tokyo | 29878 | GenBank | LC522975 |
| 20 | MN908947_Human_China_Hubei_Province_Wuhan_City | 29903 | GenBank | MN908947 |
| 21 | MN938384_Human_China_Guangdong_Shenzhen | 29838 | GenBank | MN938384 |
| 22 | MN975262_Human_China_Guangdong_Shenzhen | 29891 | GenBank | MN975262 |
| 23 | MN985325_Human_USA_Washington_Snohomish_County | 29882 | GenBank | MN985325 |
| 24 | MN988668_Human_China_Hubei_Wuhan | 29881 | GenBank | MN988668 |
| 25 | MN988669_Human_China_Hubei_Wuhan | 29881 | GenBank | MN988669 |
| 26 | MN988713_Human_USA_Illinois__Chicago | 29882 | GenBank | MN988713 |
| 27 | MN994467_Human_USA_California_Los_Angeles | 29882 | GenBank | MN994467 |
| 28 | MN994468_Human_USA_California_Orange_County | 29883 | GenBank | MN994468 |
| 29 | MN997409_Human_USA_Arizona_Phoenix | 29882 | GenBank | MN997409 |
| 30 | MT007544_Human_Australia_Victoria_Clayton | 29893 | GenBank | MT007544 |
| 31 | MT020781_Human_Finland_Lapland | 29847 | GenBank | MT020781 |
| 32 | MT020880_Human_USA_Washington | 29882 | GenBank | MT020880 |
| 33 | MT020881_Human_USA_Washington | 29882 | GenBank | MT020881 |
| 34 | MT027062_Human_USA_California | 29882 | GenBank | MT027062 |
| 35 | MT027063_Human_USA_California | 29882 | GenBank | MT027063 |
| 36 | MT027064_Human_USA_California | 29882 | GenBank | MT027064 |
| 37 | MT039887_Human_USA_Wisconsin | 29879 | GenBank | MT039887 |
| 38 | MT039890_Human_South_Korea | 29903 | GenBank | MT039890 |
| 39 | MT044257_Human_USA_Illinois | 29882 | GenBank | MT044257 |
| 40 | MT044258_Human_USA_California | 29858 | GenBank | MT044258 |
| 41 | MT049951_Human_China | 29903 | GenBank | MT049951 |
| 42 | MT066175_Human_China_Taiwan_Taipei | 29870 | GenBank | MT066175 |
| 43 | MT066176_Human_China_Taiwan_Taipei | 29870 | GenBank | MT066176 |
| 44 | NMDC60013002-06_Human_China_Hubei_Province_Wuhan_City | 29891 | NMDC | NMDC60013002-06 |
| 45 | NMDC60013002-07_Human_China_Hubei_Province_Wuhan_City | 29890 | NMDC | NMDC60013002-07 |
| 46 | NMDC60013002-08_Human_China_Hubei_Province_Wuhan_City | 29891 | NMDC | NMDC60013002-08 |
| 47 | NMDC60013002-09_Human_China_Hubei_Province_Wuhan_City | 29896 | NMDC | NMDC60013002-09 |
| 48 | NMDC60013002-10_Human_China_Hubei_Province_Wuhan_City | 29891 | NMDC | NMDC60013002-10 |
| 49 | NMDC60013002-05_Human_China_HubeiProvince_WuhanCity | 26973 | NMDC | NMDC60013002-05 |

**Table 2**. Names and Length of bat coronavirus genome sequences used in this study.

| # | Genome Sequences | Length (nt) | Database | Acc. Number |
|---|---|---|---|---|
| 1 | GWHABKP00000000_Rhinolophusaffinis_China_Yunnan_Pu'er | 29855 | Genome Warehouse | GWHABKP00000000 |
| 2 | MG772933.1_Bat_SARS-like_coronavirus_isolate_bat-SL-CoVZC45 | 29802 | GenBank | MG772933 |

**Table 3**. Synonymous and non-synonymous substitution in the SARS-CoV-2 genome.

| Ref. name | Ref. position | Ref. Base | Alt. Base | Ref. codon | Alt. codon | Group | Ref. AA | Alt. AA |
|---|---|---|---|---|---|---|---|---|
| MN908947 | 614 | G | A | GCT | ACT | 1st | Ala (A) | Thr (T) |
| MN908947 | 2662 | C | T | TAC | TAT | 3rd | Tyr (Y) | Tyr (Y) |
| MN908947 | 5084 | A | G | ATA | GTA | 1st | Ile (I) | Val (V) |
| MN908947 | 8782 | C | T | AGC | AGT | 3rd | Ser (S) | Ser (S) |
| MN908947 | 18060 | C | T | CTC | CTT | 3rd | Leu (L) | Leu (L) |
| MN908947 | 24034 | C | T | AAC | AAT | 3rd | Asn (N) | Asn (N) |
| MN908947 | 24325 | A | G | AAA | AAG | 3rd | Lys (K) | Lys (K) |
| MN908947 | 26144 | G | T | GGT | GTT | 2nd | Gly (G) | Val (V) |
| MN908947 | 26729 | T | C | GCT | GCC | 3rd | Ala (A) | Ala (A) |
| MN908947 | 28077 | G | C | GTG | CTG | 1st | Val (V) | Leu (L) |
| MN908947 | 28144 | T | C | TTA | TCA | 2nd | Leu (L) | Ser (S) |
| MN908947 | 28854 | C | T | TCA | TTA | 2nd | Ser (S) | Leu (L) |
| MN908947 | 29095 | C | T | TTC | TTT | 3rd | Phe (F) | Phe (F) |

**Table 4**. Mixed bases in the MN988713-Human-USA-Illinois-Chicago sequence.

| Genome Sequence | Nucleotides | Group |
|---|---|---|
| MN908947_Human_China_Hubei_Province_Wuhan_City | TCCCTGTC | CTC |
| MN988713_Human_USA_Illinois__Chicago | WYYYYSYY | CTC/TCC |
| MN994467_Human_USA_California_Los_Angeles | TCTTCCCC | TCC |
| MT044257_Human_USA_Illinois | ATTTCCCC | TCC |

※W means T or A, Y means C or T, S means G or C.

```
                                                                                1111111111112222222222222222222222222
                                                          1122233356667788889999111255577889000111122244556667888899
                                                        465906917700990803790455057135603050669136723037813740127803
                                                        914196777983691608883936856129007616773314020327145297459590
                                                        048212178241866618274141374547703025096764743455044937432453
--------------------------------------------------------------------------------------------------------------------
REF_genome_MN908947_Wuhan-Hu-1                          TGGCCCGCACACCTGGAACTACCCGGTCCTTCCCCTGGCAGTCCTCAGCGTTCGTCACCC
CNA0007332_Human_China_HubeiProvince_WuhanCity          ............A.............A.................................
CNA0007334_Human_China_HubeiProvince_WuhanCity          ............................................................
GWHABKF00000000_Human_China_HubeiProvince_WuhanCity     ........G........G.A........................................
GWHABKG00000000_Human_China_HubeiProvince_WuhanCity     ............................................................
GWHABKH00000000_Human_China_HubeiProvince_WuhanCity     .............C..............................................
GWHABKI00000000_Human_China_HubeiProvince_WuhanCity     ............................................................
GWHABKJ00000000_Human_China_HubeiProvince_WuhanCity     ...............T............................................
GWHABKL00000000_Human_China_Hubei_Wuhan                 ............................................................
GWHABKM00000000_Human_China_Hubei_Wuhan                 ..............A........................G....................
GWHABKN00000000_Human_China_Hubei_Wuhan                 ............................................................
GWHABKO00000000_Human_China_Hubei_Wuhan                 ................C.....T.....................................
GWHABKS00000000_Human_China_Zhejiang_Hangzhou           ............................................................
LC521925_Human_Japan_Aichi                              ...T..............................T.........................
LC522972_Human_Japan_Kyoto                              .........................T..T...................G..........T
MN988668_Human_China_Hubei_Wuhan                        ............................................................
MN988669_Human_China_Hubei_Wuhan                        ............................................................
MT027064_Human_UnitedStates_California                  ....T.....................................T.................
MT039887_Human_UnitedStates_Wisconsin                   ................................T...........................
MT044258_Human_UnitedStates_California                  ............................................................
MT066176_Human_China_Taiwan_Taipei                      ....................GT......................................
NMDC60013002-07_Human_China_HubeiProvince_WuhanCity     ............................................................
NMDC60013002-08_Human_China_HubeiProvince_WuhanCity     ............................................................
NMDC60013002-09_Human_China_HubeiProvince_WuhanCity     ....................................................T..T....
NMDC60013002-10_Human_China_HubeiProvince_WuhanCity     ....................................AA......................
GWHABKK00000000_Human_China_Hubei_Wuhan                 ........................................A.....G.............
NMDC60013002-06_Human_China_HubeiProvince_WuhanCity     ..............................................G.............
MN994468_Human_UnitedStates_California_OrangeCounty     ...............................T.................T..........
MT007544_Human_Australia_Victoria_Clayton               ...................................C........G....T..........
MT039890_Human_SouthKorea                               ......T....T...............T.C........T....G...T.TA.........
MT072688_Human_Nepal_Kathmandu                          .............................................T..............
MT027062_Human_UnitedStates_California                  .A........G..............................................T..
MT027063_Human_UnitedStates_California                  .A........G..............................................T..
CNA0007335_Human_China_HubeiProvince_WuhanCity          ..................T...................................C.....
MT049951_Human_China                                    ..................T.....C................A............C.....
MT066175_Human_China_Taiwan_Taipei                      ..................T...................................C.....
MN985325_Human_UnitedStates_Washington_SnohomishCounty  ..................T..............T....................C.....
MT020880_Human_UnitedStates_Washington                  ..................T..............T....................C.....
MT020881_Human_UnitedStates_Washington                  ..................T..............T....................C.....
MN994467_Human_UnitedStates_California_LosAngeles       ..A...............T..........................T.....C.CC.T...
MT044257_Human_UnitedStates_Illinois                    A......T..........T..........................T.....C.CC.....
MN938384_Human_China_Guangdong_Shenzhen                 ..................T...................................C...T.
MN975262_Human_China_Guangdong_Shenzhen                 ..................T....T......C.......................C...T.
MN997409_Human_UnitedStates_Arizona_Phoenix             ..................T.....T.............................C...T.
LC522973_Human_Japan_Tokyo                              .....T...T........T...................................C...T.
LC522974_Human_Japan_Tokyo                              .....T............T...................................C...T.
LC522975_Human_Japan_Tokyo                              .....T............T...................................C...T.
```

**Figure S1. Variable sites found from SARS-CoV-2 ORF sequences**

After alignment, the SARS-CoV-2 sequences had 60 variable sites from 29,145 bps alignment of ORF sequences from 47 samples isolated from human. Dots indicate the same as the reference base on the top of the alignment. A, C, G or T indicate the difference in the reference base. Nucleotide position numbers in the reference genome are shown above the alignment. All sites including gaps were removed from the analysis.

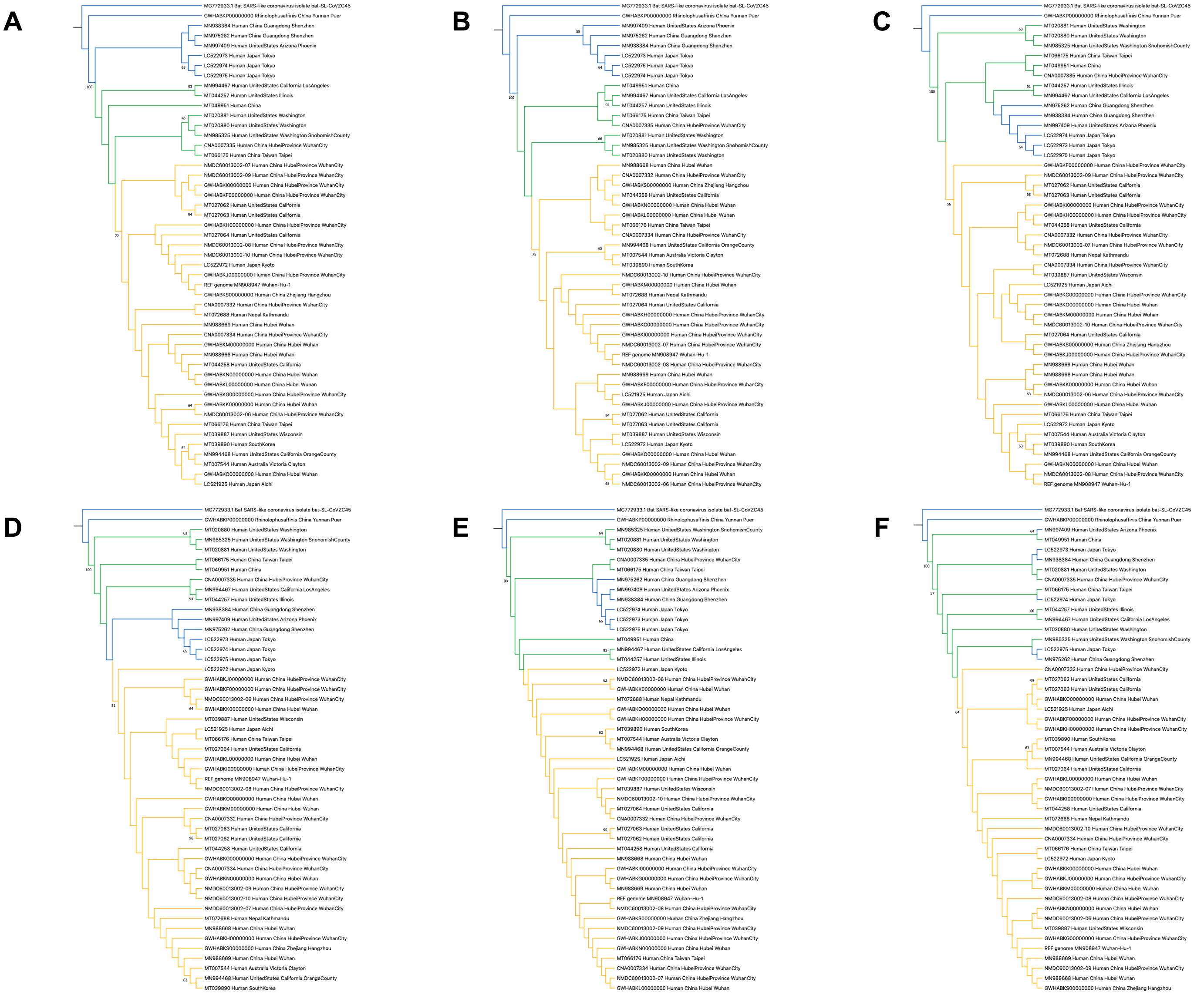

**Figure S2. Topological phylogeny of SARS-CoV-2**

Maximum-likelihood phylogenetic analyses under each condition below. (A) DNA-based no separate condition. Final log-likelihood was -58648.257750. AIC score was 117504.515500. (B) DNA-based separating 1st/2nd and 3rd codon position condition. Final log-likelihood was -56210.666246. AIC score was 111828.725349. (C) DNA-based separating each gene condition. Final log-likelihood was -59189.786542. AIC score was 118667.573085. (D) DNA-based separating each gene and also 1st/2nd and 3rd codon position condition. Final log-likelihood was -56528.571686. AIC score was 113415.143371. (E) DNA-based separating each gene and each codon position condition. Final log-likelihood was -56171.539079. AIC score was 112759.078158. (F) Protein-based separating each gene condition. Final log-likelihood was -34553.898182. AIC score was 69363.796363. Branch length was meaningless.